\documentclass[aip,rsi,reprint,graphicx,floatfix]{revtex4-1} 
\usepackage{graphicx}
\usepackage{float}
\usepackage[colorlinks=true,citecolor=blue,urlcolor=blue,hypertexnames=true]{hyperref}

\let\oldmaketitle\maketitle
\renewcommand\maketitle{{\bfseries\boldmath\oldmaketitle}}

\begin{document}


\title{The new versatile general purpose surface-muon instrument (GPS) based on silicon photomultipliers
for 
${\mu}$SR measurements on a continuous-wave beam} 



\author{A. Amato}
\email[]{alex.amato@psi.ch}
\affiliation{Laboratory for Muon-Spin Spectroscopy, Paul Scherrer Institute, 5232 Villigen PSI, Switzerland}
\author{H. Luetkens}
\affiliation{Laboratory for Muon-Spin Spectroscopy, Paul Scherrer Institute, 5232 Villigen PSI, Switzerland}
\author{K. Sedlak}
\affiliation{Swiss Plasma Center, Ecole Polytechnique F\'ed\'erale de Lausanne, 5232 Villigen PSI, Switzerland}
\author{A. Stoykov}
\affiliation{Laboratory for Particle Physics, Paul Scherrer Institute, 5232 Villigen PSI, Switzerland}
\author{R. Scheuermann}
\affiliation{Laboratory for Muon-Spin Spectroscopy, Paul Scherrer Institute, 5232 Villigen PSI, Switzerland}
\author{M. Elender}
\affiliation{Laboratory for Muon-Spin Spectroscopy, Paul Scherrer Institute, 5232 Villigen PSI, Switzerland}
\author{A. Raselli}
\affiliation{Laboratory for Scientific Developments and Novel Materials, Paul Scherrer Institute, 5232 Villigen PSI, Switzerland}
\author{D. Graf}
\affiliation{Laboratory for Scientific Developments and Novel Materials, Paul Scherrer Institute, 5232 Villigen PSI, Switzerland}


\date{\today}

\begin{abstract}
\rightskip.5in

We report on the design and commissioning of a new spectrometer for muon-spin relaxation/rotation studies installed at the Swiss Muon Source (S$\mu$S) of the Paul Scherrer Institute (PSI, Switzerland). This new instrument is essentially a new design and replaces the old general-purpose surface-muon instrument (GPS) which has been for long the workhorse of the $\mu$SR user facility at PSI. By making use of muon and positron detectors made of plastic scintillators read out by silicon photomultipliers (SiPMs),
a time resolution of the complete instrument of about 160~ps (standard deviation) could be achieved. In addition, the absence of light guides, which are needed in traditionally built $\mu$SR instrument to deliver the scintillation light to photomultiplier tubes located outside magnetic fields applied, allowed us to design a compact instrument with a detector set covering an increased solid angle compared to the old GPS.     

\end{abstract}


\maketitle 

%
\section{Introduction}
\subsection{General Introduction}
The muon-spin relaxation/rotation technique (hereafter labeled $\mu$SR) uses implanted muons (usually positive) to probe properties of condensed matter and molecular systems at the atomic level (see for example Ref. \onlinecite{Schenck, Amato, Blundell, Yaouanc, Turner}). 
The technique is used to investigate magnetic systems, spin dynamics, superconducting states, chemical radicals or hydrogen
behavior in semiconductors.

A polarized-muon beam is obtained by collecting the muons produced via the two-body decay of positive pions $\pi^+ \rightarrow \mu^+ + \nu_{\mu}$. The pions themselves are created in the production targets placed in a high-energy proton accelerator (usually about 600~MeV). As the pion has zero spin and only left-handed neutrinos $\nu_{\mu}$ exist, the decay muons have the spin antiparallel to their momentum in the pion rest
frame.
At S$\mu$S, muon beams of different energy ranges are available\cite{homepage_PSI}: 
i) a beam of high-energy muons ($\sim$40-50~MeV) is routinely used to study specimens contained in controlled environments, e.g. samples inside pressure cells; 
ii) unique to S$\mu$S, a muon beam of tunable very low energy (down to the eV-keV range), corresponding to implantation depths in solids from a few nanometers up to several hundred nanometers, allowing one to perform thin-film, near-surface and multi-layer studies; 
iii) and finally the majority of the muon beams at PSI make use of muons produced still inside, but near
the surface, of the production target (i.e. from pions decaying at rest). 
For the latter beam (often called ``surface'' or ``Arizona'' beam\cite{arizona}) the muons are 100\% polarized and ideally monochromatic, having a very low momentum of 29.8~MeV/c corresponding to a kinetic energy of 4.1 MeV. Implanted in matter, they have a stopping range of the order of 0.180~g/cm$^2$. 

Making use of this latter type of beam, the old general purpose surface-muon instrument (GPS) located on the $\pi$M3.2 beamline at S$\mu$S has been the workhorse of the PSI $\mu$SR user facility, as reflected by about 40 peer-reviewed publications per year \cite{homepage_PSI}, but relied on technology choices made more than 25 ago.

The present paper describes the design and commissioning of an improved general purpose surface-muon instrument. The defined goals were to improve the time resolution of the spectrometer, to increase the solid angle coverage of the positron detectors and to possibly increase the available magnetic field produced by normal-conducting Helmholtz coils.

\begin{center}
\begin{figure*}[t]
\includegraphics[width=0.7\textwidth]{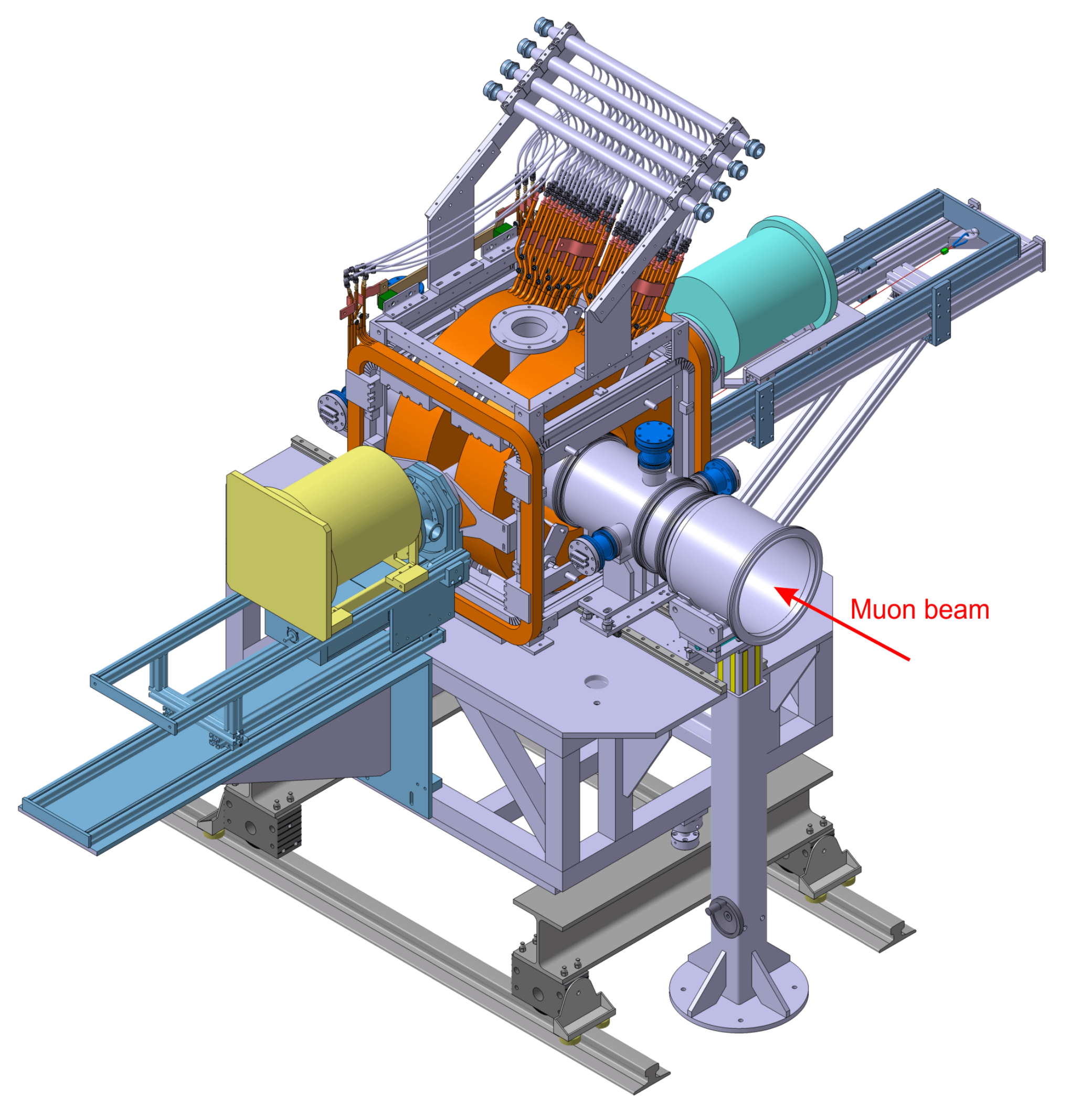}
\caption{Schematics of the new GPS instrument located on the $\pi$M3.2 beamline of the HIPA complex of the Paul Scherrer Institute. The instrument is directly connected with the vacuum pipe of the beamline. The yellow and green elements represent cryostats located on the so-called first and second cryogenics port, respectively. The main Helmholtz coils as well as the coils of the auxiliary magnet are shown in orange. An idea of the dimensions is provided by the outer diameter of one of the main magnet coils which is 742~mm.}
\label{New_GPS_view}
\end{figure*}
\end{center}
\subsection{Short Introduction to the $\mu$SR Technique}
In this section a brief introduction to the $\mu$SR technique is provided. Interested readers are referred to more comprehensive textbooks  and review articles \cite{Schenck, Amato, Blundell, Yaouanc, Amato_2}.

If the $\mu^+$ implanted into the sample are subject to magnetic interactions, their polarization becomes time dependent and is noted $\mathbf{P}_{\mu}(t)$.
The radioactive decay of the muons is described by $e^{-(t/\tau_{\mu})}$, where $t$ is the decay time after the implantation and $\tau_{\mu}\simeq2.197 \times 10^{-6}$~s is the muon lifetime. The $\mu$SR techniques is based on the weak decay of the muon 
($\mu^+\rightarrow e^+ + \nu_e + \bar{\nu}_{\mu}$) which, because of the parity violation in the weak interaction, produces an asymmetric distribution of the emitted positron with respect to the muon spin direction at the decay time. Therefore, by measuring for a large enough muon ensemble the spatial positron distribution as a function of the muon decay time in different detectors located around the sample it becomes possible to determine the time evolution of the muon polarization $\mathbf{P}_{\mu}(t)$.

For a positron detector $i$ placed in  direction $\mathbf{\hat{n}}_i$ with respect to the initial muon polarization $\mathbf{P}_{\mu}(0)$, the time histogram of the collected time intervals between the muon implantation and the positron detection has the form
\begin{equation}
 \label{equation_general}
  N_i(t) = B_i + {N_0}_i
  \exp(-\frac{\displaystyle t}{\displaystyle\tau_{\mu}}){\Big [}1 + {a_i}\,\mathbf{P}_{\mu}(t)\cdot\mathbf{\hat{n}}_i{\Big ]}~,
\end{equation}
where $B_i$ is a time-independent background, ${N_0}_i$ is a normalization constant and the exponential accounts for the muon decay. 
The asymmetry of this detector is then given by ${A_i(t)} = {a_i}\,\mathbf{P}_{\mu}(t)\cdot\mathbf{\hat{n}}_i$, with $P_{\mu}(0) = |\mathbf{P}_{\mu}(0)|$ corresponding to the beam polarization (of the order of $\sim$1). The parameter $a_i$ is determined, on one side, by the intrinsic asymmetry of the weak decay mechanism and, on the other side, by the detector solid angle, the detector efficiency and also by the absorption and scattering of positrons in materials located on the path between the sample and the detector. An average of  $\langle a_i\rangle = 1/3$ is obtained if all emitted positrons are detected with the same efficiency irrespective of their energy by an ideally small detector. Practically, and mainly due to the finite solid angle of the detectors, typical values of ${a_i}$  lie between 0.25 and 0.30.

The time dependence of the muon polarization function $\mathbf{P}_{\mu}(t)$ depends on the average value, distribution and time evolution of the internal fields and hence contains all the relevant physics of the magnetic interactions of the muon with its surrounding inside the sample. 
${A_i(t)}$ is often labeled as  the $\mu$SR signal, whereas its envelope is usually referred as the $\mu^+$--depolarization function. Since the other parameters included in Eq.~\ref{equation_general} do not contain physical information about the magnetic properties of the sample investigated, usually solely the $\mu$SR signal ${A_i(t)}$ is reported as a result of $\mu$SR measurements.

The $\mu$SR signal can be extracted by fitting each individual histogram using Eq.~\ref{equation_general} or, in particular in measurements performed in zero field (ZF) or longitudinal field (LF) configurations, it can be directly obtained from two positron counters mounted on the opposite sides of the sample in the Forward (F) and Backward (B) detectors with respect to the beam direction. In the case of an instrument making use of a surface beam (like GPS), for the ZF and LF configurations the initial muon-spin direction is essentially antiparallel to the beam direction. Note that a slight angle ($\theta < 10^\circ$) arises due to the muon-spin rotation in the field of the spin-rotator device used here as a Wien filter to clean the incident muon beam from the positron contamination. In these configurations, one may write:
\begin{equation}
 {A}(t)= \frac{(N_{{\text B}}(t) - B_{\text B})  - \alpha (N_{{\text F}}(t)- B_{\text F})}{(N_{{\text B}}(t) - B_{\text B})  + \alpha (N_{{\text F}}(t)- B_{\text F})},
 \label{asy_formula}
\end{equation}
The parameter $\alpha$ takes into account the efficiency and the different solid angles of both positron detectors. As the latter ones are also dependent on the exact shape of the sample, the parameter $\alpha$ must always be calibrated at the beginning of the experiment (see also Section~\ref{alpha_in field}). This calibration is usually performed by applying, in the paramagnetic phase of the sample, a small field (typically of the order of 5~mT) perpendicular to the muon beam axis (in the case of GPS this field is applied with the auxiliary magnet, see Fig.~\ref{New_GPS_view}) and by fitting the Eq.~\ref{asy_formula} to the observed $\mu$SR signal.  

Note that for some experiments the direction of the initial muon-spin direction needs to be ideally perpendicular to the muon-beam direction. This is mandatory when performing so-called transverse-field (TF) experiments with a large magnetic field (more than 10~mT; e.g. for Knight-shift measurements or to  determine the field distribution in the Abrikosov state of a type II superconductor). As such fields have to be applied along the beam axis to avoid a large beam deflection due to the relatively small muon-momentum value, the muon-spin needs to be turned as much as possible in the perpendicular direction by the spin-rotator (i.e. $\theta \rightarrow 90^\circ$) \cite{Beveridge}. Here again the $\mu$SR signal can be extracted by fitting each individual histogram (see Eq.~\ref{equation_general}) or can be obtained using Eq.~\ref{asy_formula} from two positron counters mounted on the opposite sides of the initial muon-spin direction. 

Pulsed muon beams, as the ones available at ISIS\cite{Hillier} (UK) and J-PARC\cite{Miyake} (Japan), present the advantages of very high muon rates and low background. Whereas a muon detector (hereafter called: M-detector), detecting the exact muon arrival time, is not needed for such beams, the high rate can only be exploited by a very high granularity of the positron detectors in order to master the large instantaneous positron rate occurring just after the muon bunch implantation (see for example Ref.~\onlinecite{Lord}). On the one hand-side pulsed beams are very well suited to investigate specimens with very weak magnetism and/or dynamics in the paramagnetic state, but on the other hand-side, due to the final width -- of the order of 80~ns -- of the incoming muon pulse, pulsed-beam $\mu$SR instruments have a reduced time resolution which on one-side limits TF $\mu$SR measurements to low applied field (of the order of 0.05~T) and on the other side impedes the determination of high spontaneous muon-spin frequencies in magnetic systems. 

At the opposite, $\mu$SR instruments on continuous wave (cw) muon beams, as the ones available at TRIUMF\cite{TRIUMF,Brewer} (Canada) and S$\mu$S, require an M-detector. In addition the data rate is limited to 
be able to correlate a muon implanted into the sample with the corresponding positron arising from its decay. Beside this limitation, cw muon-beam $\mu$SR instruments present several advantages compared to pulsed-beam instruments: i) the number of detectors can be drastically reduced due to the absence of the  problems inherent to the instantaneous positron rate in pulsed beams; ii) the size of the incoming beam can be very efficiently reduced using an active veto detector (see Section~\ref{detectors_geometry}) removing essentially any background signals; iii) the superior time resolution resulting from the detection of individual muons.
%
%
\section{Muon and Positron Detectors}
\subsection{Detector Geometry}
\label{detectors_geometry}
The route chosen to improve the GPS detector geometry is based on the use of so-called silicon photomultipliers (SiPM). SiPM based $\mu$SR detectors have been first developed to equip the avoided level-crossing instrument (ALC) at PSI\cite{Stoykov_1,Sedlak_1}. This was the first step and a proof of principle for the consequent development of SiPM-based detectors for the new high magnetic field instrument HAL-9500 at PSI \cite{Sedlak_2,Stoykov_2,HAL}. These solid-state photodetectors have been shown to deliver performances similar to that of photomultiplier tubes (PMT) but present several advantages as their compactness and the fact that their performance is insensitive to magnetic fields. This last point allows one to mount the photodiodes directly on the scintillator part of the detector. Hence, one may completely get rid of the Plexiglas light-guides indispensable when working with PMTs in magnetic-field environment, which were: i) introducing an attenuation and broadening of the light pulses and therefore limiting the overall time resolution of the detectors; and ii) occupying valuable volume space and therefore putting severe geometrical constraints for the detectors geometry and sample environment.

\begin{figure*}[t]
 \begin{minipage}{0.58\textwidth}
 \includegraphics[width=\linewidth]{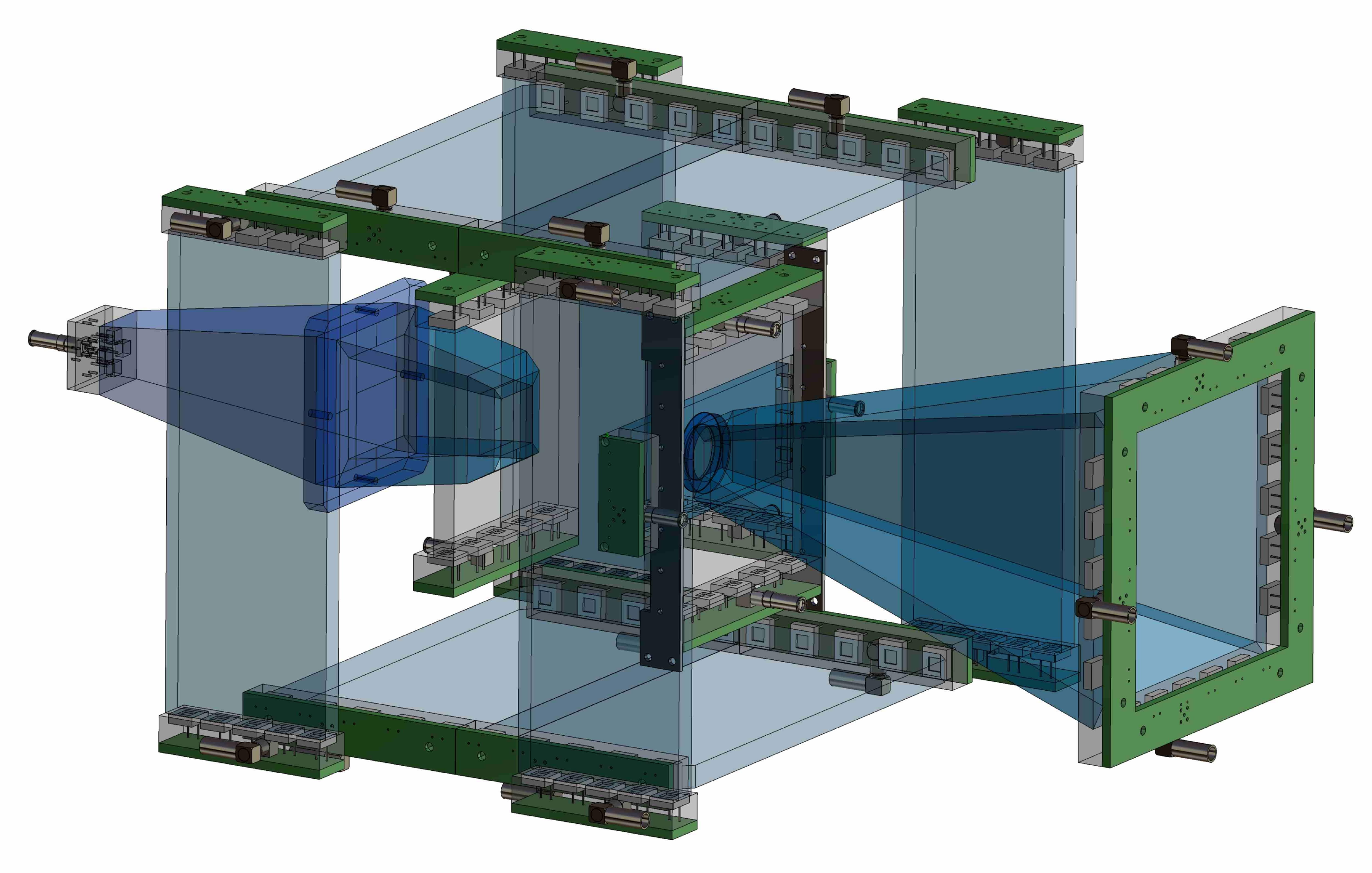}
 \end{minipage}
 \begin{minipage}{0.4\textwidth}
 \includegraphics[width=\linewidth]{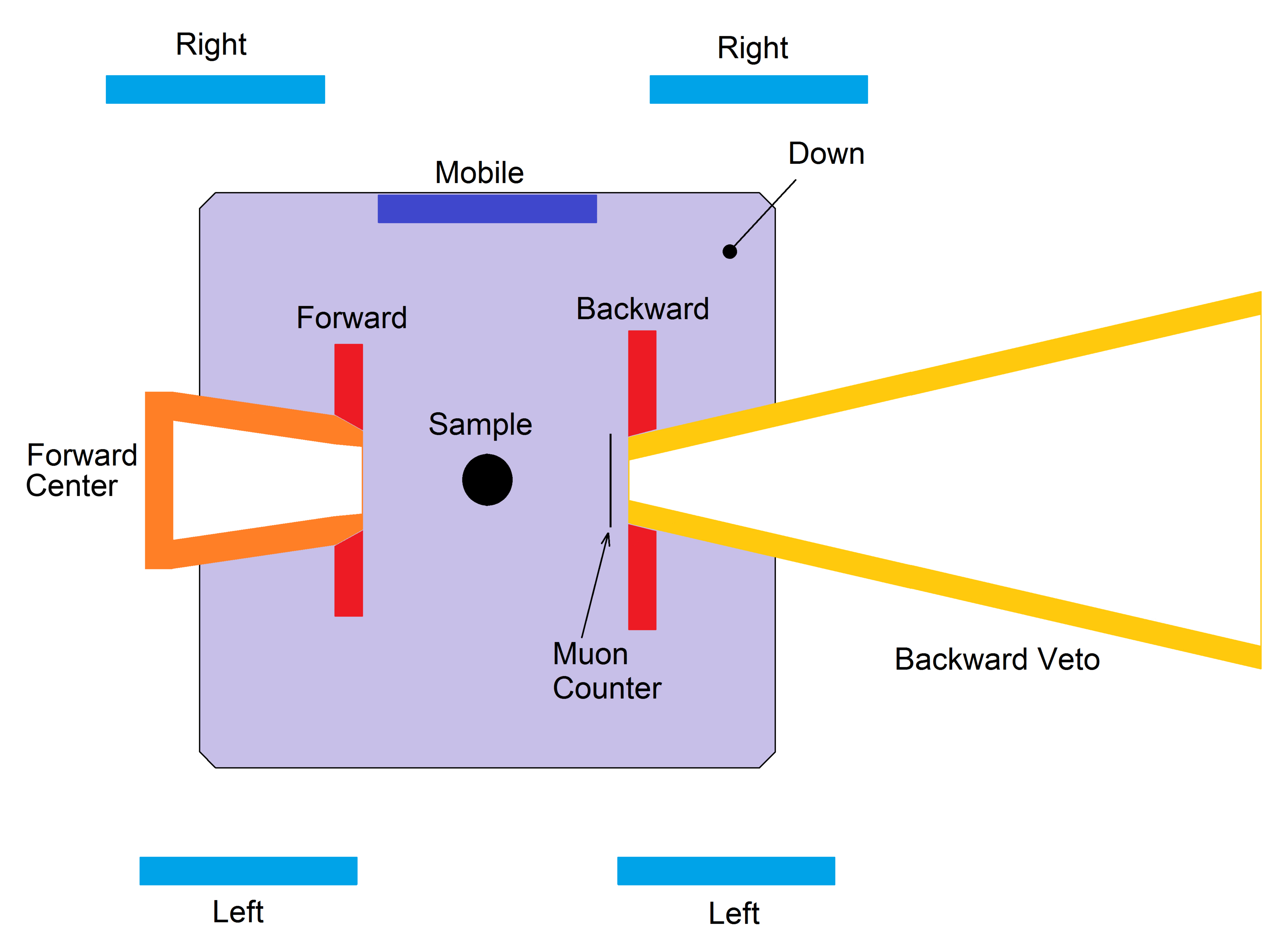}
 \end{minipage}
\caption{Left: 3D-view of the muon and positron detectors of the GPS instrument where the SiPM-readout is visible. Right: Cross-section at the sample level (top view). The muons are entering the spectrometer from the right-hand side, i.e. parallel to the axis of the Backward veto pyramid. The lateral dimension of the Down detector is 10~cm.}
\label{figure_detectors}
\end{figure*}

Figure~\ref{figure_detectors} exhibits the layout of the detector system of the GPS instrument, which is directly mounted in the vacuum beam pipe around the cryogenics equipment inside which the sample is located. The detector system is made of timing and Veto detectors. The timing detectors consist of the M-detector and the positron detectors labeled Forward, Backward, Up (U), Down (D), Right (R) and Left (L). Note that as above the denomination of the detectors is given with respect to the beam direction.  
The veto detectors are set to detect a muon and its decay positron and are constituted by the backward ``pyramid'' $\text{B}_{\text {veto}}$ and the forward veto. The backward veto pyramide, which acts as an active collimator, has its principal axis along the beam path and a square opening at its tip of  $7\times 7$~mm$^2$ where the M-detector is located.
For small samples (i.e. smaller than the beam width of diameter 10.5~mm at the sample position) the central part of the F detector 
($\text{F}_{\text{center}}$) is also used  as part of the veto detectors. This is made possible as along the beam path, between the sample and the forward detectors solely minimum material is present (i.e. a $10\,\,\mu$m thick titanium window and two layers of superisolation of 12$\,\,\mu$m thickness) thus making sure that muons missing the sample will end-up in the $\text{F}_{\text{center}}$ detector. When measuring bulky samples or samples mounted on a thick and large sample holder $\text{F}_{\text {center}}$ is added to the definition of the F positron detector (see Table~\ref{table_logic}).

A mobile detector ($\text{P}_{\text {mob}}$) is mounted rigidly on rods connecting the two so-called ``cryogenics ports'' located on an axis horizontal and perpendicular to the beam direction. Therefore, depending on the cryogenics port used and brought on the beam axis, the $\text{P}_{\text {mob}}$ will be added to the definition of the R detector when  using the first crogenics port (i.e. $\text{R}= \text{R}_{\text {back}}+\text{R}_{\text {forw}}+\text{P}_{\text {mob}}$ and $\text{L}= \text{L}_{\text {back}}+\text{L}_{\text {forw}}$)  or the L detector when  using the second cryogenics port (i.e. $\text{L}= \text{L}_{\text {back}}+\text{L}_{\text {forw}}+\text{P}_{\text {mob}}$ and $\text{R}= \text{R}_{\text {back}}+\text{R}_{\text {forw}}$). Note that to simplify the mounting, the detectors U and D are also physically split between a forward and a backward part.

\begin{table*}[tb]
\begin{tabular*}{\textwidth}{l@{\extracolsep{\fill}}lll}\hline\hline
Incoming Muon& & $\text{M}\cdot\overline{\text{B}_{\text {veto}}}$&\\
Muon event (``Stopped Muon'')& &$\text{M}\cdot\overline{\text{V}}$&\\
Positron event & &$\text{P}\cdot\overline{\text{V}}$ & \\
\hline
&``No-Veto'' mode& &``Veto'' mode\\
&\textit{Large Samples}&&\textit{Small Samples}\\
Veto definition & $\text{V}=\text{B}_{\text {veto}}$ & &$\text{V}=\text{B}_{\text {veto}}+\text{F}_{\text {center}}$\\
Forward definition & $\text{F}= \text{F}_{\text {out}}+\text{F}_{\text {center}}$ & &$\text{F}= \text{F}_{\text {out}}$\\
\hline\hline
\end{tabular*}
\caption{Definitions of the Veto and Forward detectors when measuring large (``No-Veto'' mode) and small (``Veto'' mode) samples. In both cases a muon event is defined as $\text{M}\cdot\overline{\text{V}}$ and a positron event as $\text{P}\cdot\overline{\text{V}}$ where P represents one of the positron detectors.}
\label{table_logic}
\end{table*}The efficiency of the full detector system has been simulated using the software applications {\it musrSim} and {\it musrAna}\cite{Sedlak_3}. The {\it musrSim} application is based on the {\sc Geant4} radiation transport toolkit\cite{Geant4} for Monte Carlo simulations and is tailored to the needs of the $\mu$SR community. It calculates the detectors response to the muons and their decay products and may also simulate   
the light transport in scintillators and its subsequent detection by a photosensitive detector.
Subsequently the output of {\it musrSim} is analysed with the general $\mu$SR analysis tool, {\it musrAna}. It allows one to implement the full logic of a real $\mu$SR experiment, as the coincidences and anti-coincidences between different detectors, and is able to extract the time-independent background of the detector histograms (see parameter $B_i$ of Eq.~\ref{equation_general}) which arises from uncorrelated muon-positron events.

The simulations have been used in particular to define an optimum shape of the  $\text{F}_{\text {center}}$ when used as a Veto detector (small samples, see Table~\ref{table_logic}). The particular shape is required: i) to minimize the probability that 
a positron originating from the sample will hit both the Forward detector ($\text{F}= \text{F}_{\text {out}}$) and the Veto detector $\text{F}_{\text {center}}$, which would result in an invalid positron event through the condition $\text{P}\cdot\overline{\text{V}}$; and ii) to make sure that the decay positron arising from a muon stopping in $\text{F}_{\text {center}}$ will be properly detected in the same detector and will therefore be rejected through the same condition $\text{P}\cdot\overline{\text{V}}$. An additional criteria for the shape of $\text{F}_{\text {center}}$ was the construction simplicity and to get away from the complicated ``champagne glass''-shape adopted in the old GPS instrument.

The reliability of the simulations has been tested by comparing for different parameters the measured and simulated values. The comparisons always report an excellent agreement. For example the experimental values of the asymmetry parameter $A$ for the  Forward and Backward detectors are respectively $A_{\text{F,exp}} = 0.244(2)$ and $A_{\text {B,exp}} = 0.259(2)$, whereas the simulations provide $A_{\text {F,sim}} = 0.246(5)$ and $A_{\text {B,sim}} = 0.257(5)$. All the experimentally obtained $\mu$SR time spectra were analyzed using the free software package {\sc musrfit}\cite{musrfit}.
\subsection{Detector Readout and Electronics}
The light from the scintillators is read-out on two opposite edges (with the exception on the $\text{B}_{\text {veto}}$ and $\text{F}_{\text {center}}$ where solely one edge is read-out) by arrays of 4 or 5 SiPMs mounted on a printed electronics board and glued directly onto the scintillator material. Each SiPM has an active area of $3\times 3$~mm$^2$, with a cell size of $40\times 40$~$\mu$m$^2$, and is enclosed in a plastic package (ASD-NUV3S-P from AdvanSiD S.R.L.). They are powered by a PSI-homemade multi-channel power supply PHV8\_600VLC. The analog signals from the SiPMs are first extracted from the vacuum beam pipe with special coaxial feedthroughs and amplified by broad-band amplifiers\cite{Stoykov_2,Cattaneo}.
The output of the amplifiers are processed by constant fraction discriminators (type PSI CFD-950) \cite{Thomas_muSR_2008} and the signals are finally sent to a multihit time-to-digital converter (CAEN, TDC V1190B). 
The event time for a detector is calculated from the average time between the events recorded on both opposite edges. 

The TDC is controlled by a data acquisition (DAQ) frontend based on the MIDAS\cite{Midas} software library which contains the full event logic, thus avoiding any hardware complication when modifying or extending the event logic, as for example switching from ``No-Veto'' to ``Veto'' mode and/or when switching between the cryogenics port and therefore changing the R and L definitions. 

\subsection{Detector Performance}
 \subsubsection{Veto Anticoincidences}
 \label{Vetos_Anticoincidences}
 Our first approach has been to simply overtake the settings adopted in the old GPS instrument concerning the criteria defining (anti-)coincidence events between different detectors. In the following, we describe problems encountered by such an approach and the solutions adopted.

\begin{center}
\begin{figure}[t]
\includegraphics[width=0.75\columnwidth]{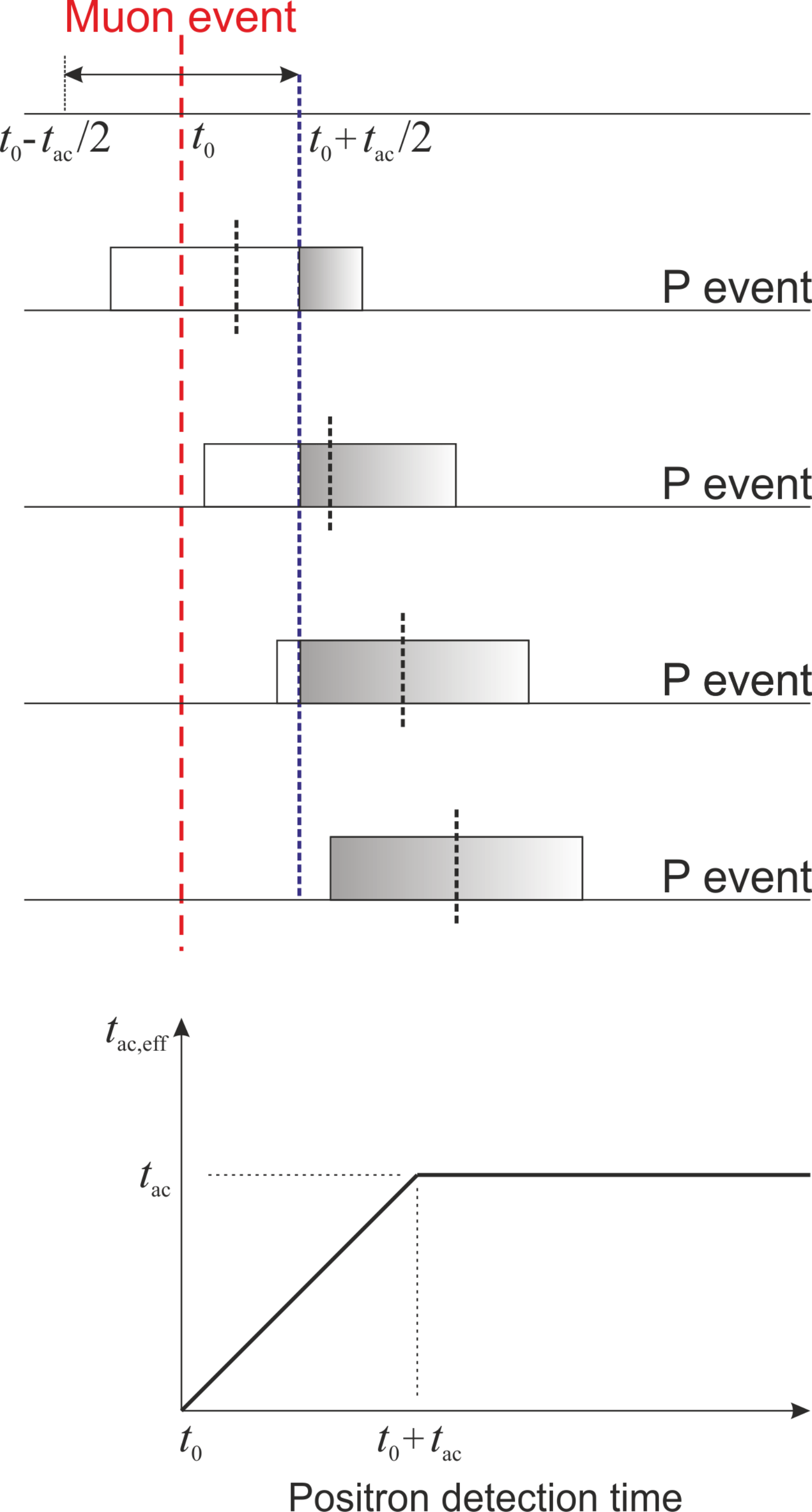}
\caption{Visualization of the decrease of the effective length $t_{\text{ac,eff}}$ of the anticoincidence time-window (represented by the grey surface on the upper panel) around the P event upon decreasing the time difference between the muon and the positron detection (see also text). The time of the muon detection is marked by the red dashed line and different examples of positron detection times are marked by the black dashed lines.}
\label{veto_logic}
\end{figure} 
\end{center}

As shown in Table~\ref{table_logic}, muon and positron events are defined to be in anticoincidence with veto events. In the DAQ frontend software, the search for anticoincidences is performed in a time-window of $\pm t_{\text{ac}}/2$ around a considered event. 
However, for very short time after the detection of a good muon at time $t_0$, the effective length $t_{\text{ac,eff}}$ of the anticoincidence time-window for the positron event (i.e. $\text{P}\cdot\overline{\text{V}}$) is decreased as a good muon event implies necessarily the absence of veto event between the times $t_0-t_{\text{ac}}/2$ and $t_0+t_{\text{ac}}/2$ (see Fig.~\ref{veto_logic} for details).

\begin{center}
\begin{figure}[t]
\includegraphics[width=0.8\columnwidth]{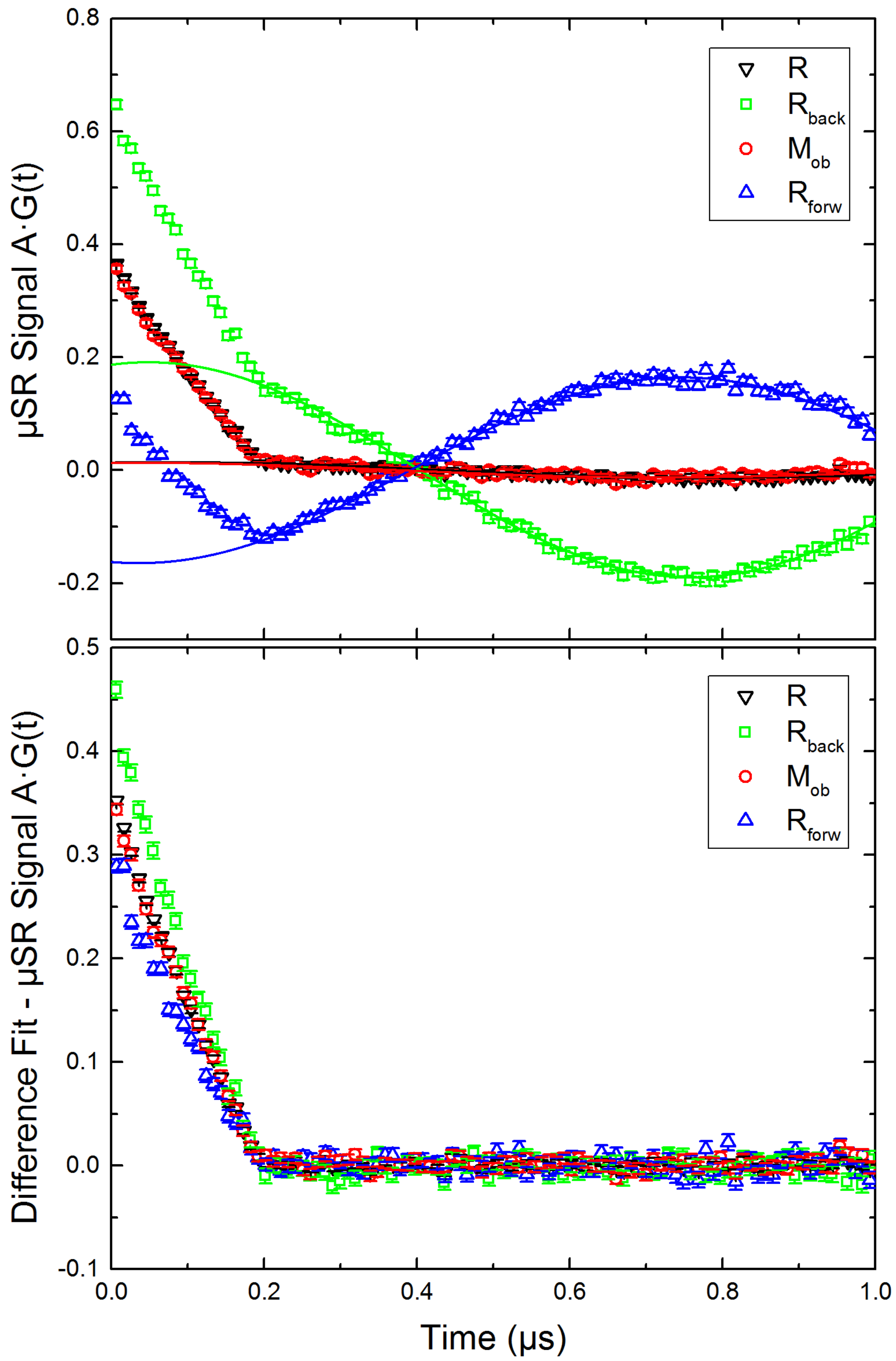}
\caption{Upper panel: Early times of the $\mu$SR signals for the different parts of the R detector obtained by fitting
Eq.~\ref{equation_general} to the raw data for a run with anticoincidence time-window increased to 200~ns. This value was specifically chosen to magnify the effect of the reduction of the effective anticoincidence time-window for short times. The lines are fits performed for times higher than 250~ns. The data were obtained on a small silver sample in an external field $\mu_0H=5$~mT applied perpendicular to the muon beam direction. The lower panel exhibits the difference between fits and the data shown in the upper panel.}
\label{veto_anti}
\end{figure} 
\end{center}

The consequence is that the number of veto events which will come into play to ``filter'' a P event will increase linearly between $t_0$ and $t_{\text{ac}}$ and will remain constant for later times. Hence, bad P events (i.e. uncorrelated ones) will have a larger probability to be registered as good P event in the time window between $t_0$ and $t_{\text{ac}}$ as examplified by the data reported on Fig.~\ref{veto_anti}, where the anticoincidence time-window was drastically increased to $t_{\text{ac}}= 200$~ns. Note that this problem will be detectable when the number of vetos events is large, i.e. when measuring very small samples, and that it will only be detectable when performing single histogram fits (i.e. using Eq.~\ref{equation_general}), whereas it will essentially cancel for asymmetry fits (using Eq.~\ref{asy_formula}). In addition, a reduction of the width of the anticoincidence time-window drastically reduces the problem. During normal operation an anticoincidence time-window of either $t_{\text{ac}}= 2.5$~ns or 5~ns has been chosen.

 \subsubsection{Coping with Veto Deadtime}
\begin{center}
\begin{figure}[t]
\includegraphics[width=0.8\columnwidth]{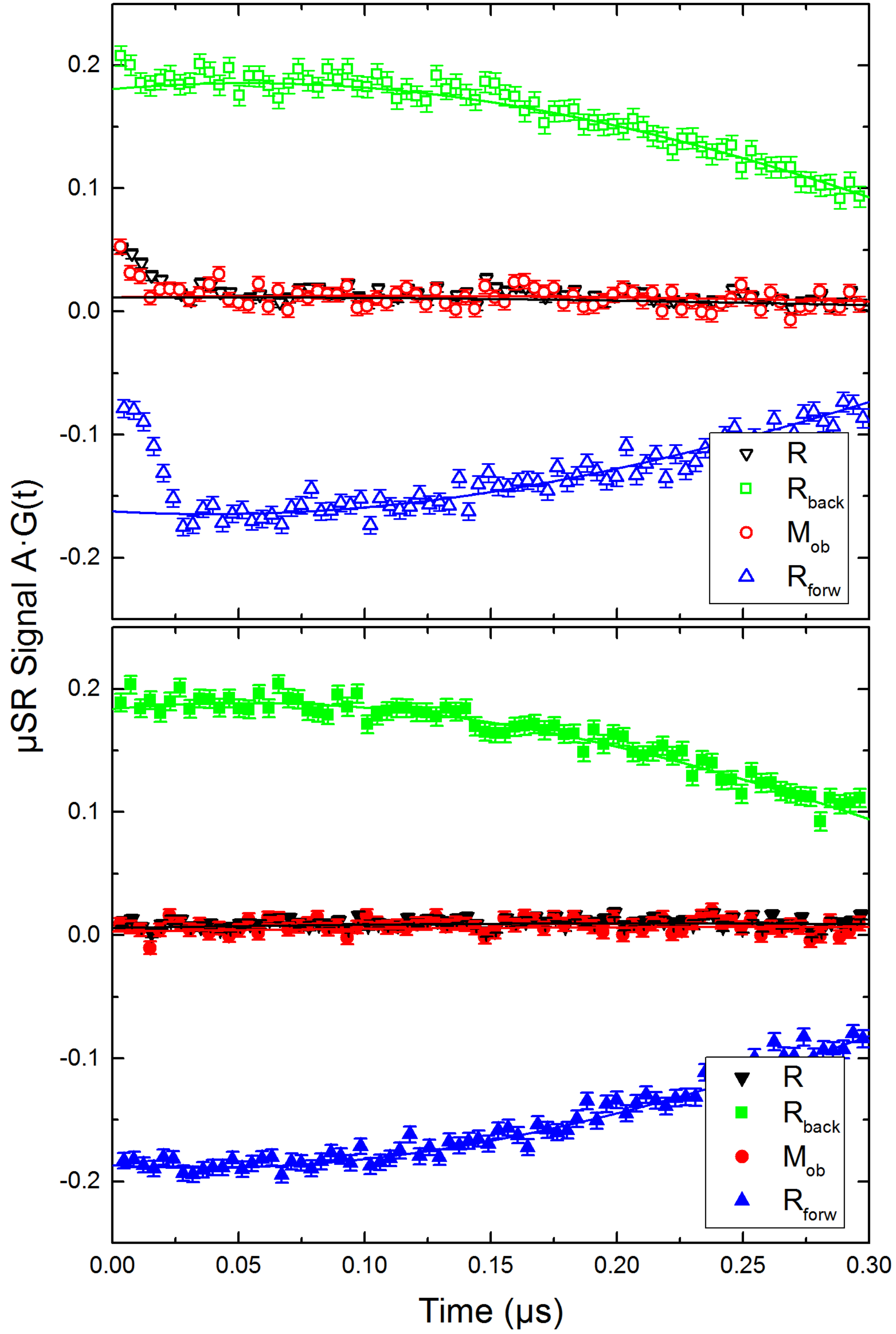}
\caption{Early times of the $\mu$SR signals for the different parts of the R detector obtained by fitting Eq.~\ref{equation_general} to the raw data. On the upper panel, data obtained with the standard symmetric veto-coincidence time-window are shown. Note that the distortion is more pronounced for detectors  presenting a larger probability to detect an event originating from $\text{F}_{\text {center}}$.
On the lower panel, data obtained with an asymmetric veto-coincidence time-window are shown (see text). All the data were obtained on a small silver sample in an external field $\mu_0H=5$~mT applied perpendicular to the muon beam direction.}
\label{deadtime}
\end{figure}
\end{center}

\begin{center}
\begin{figure}[t]
\includegraphics[width=0.8\columnwidth]{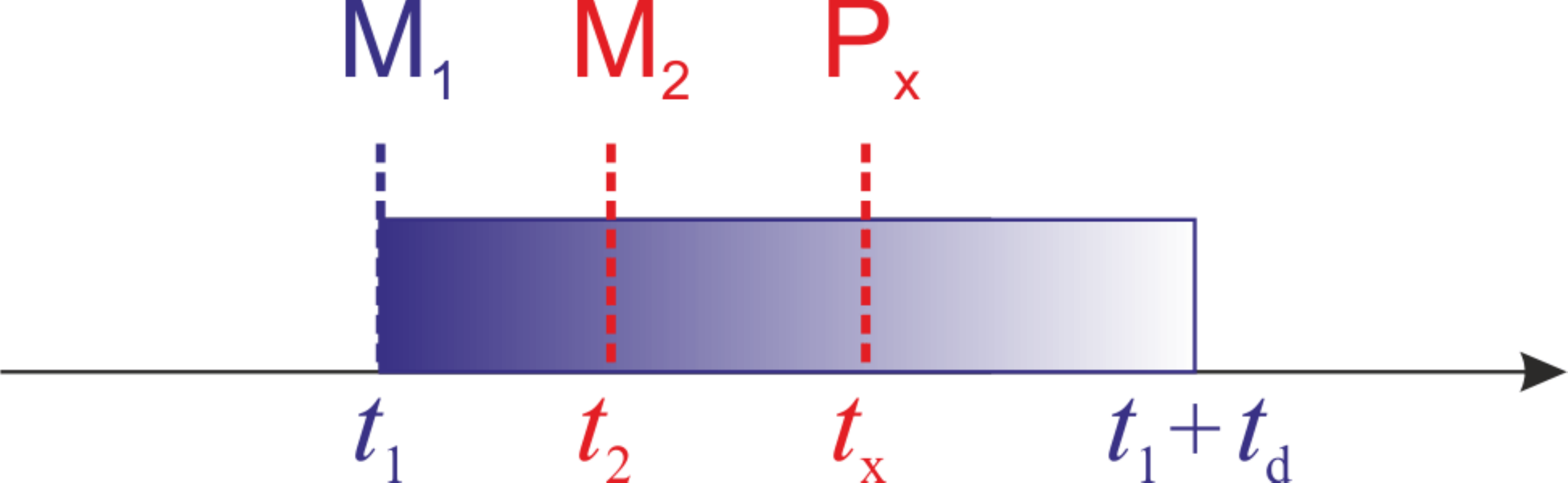}
\caption{Visualization of a possible M$_2$ -- P$_x$ pair not detected in the Veto detector $\text{F}_{\text {center}}$ as a result of the deadtime after an event M$_1$. The event M$_2$ will be considered as a good muon event, whereas P$_x$ will be considered as a good positron event if detected by a positron detector. Note that $t_{d}\gg t_{ac}$.}
\label{APD_deadtime}
\end{figure}
\end{center}

After solving the problem with the overall anticoincidence time-window described above in Section~\ref{Vetos_Anticoincidences} and by performing high statistics detector tests in the ``Veto''-mode with a large number of incoming muons, a slightly increased rate was observed on different detectors just after the prompt peak characterizing $t_0$ (see Fig.~\ref{deadtime}). The problem was identified as a dead-time of the detectors 
after a first firing. This dead-time problem is stronger in the $\text{F}_{\text {center}}$ detector due to the entire light collection on one side of the detector. As a direct consequence, other detectors presenting a large probability to detect an event originating from $\text{F}_{\text {center}}$ (such as the decay positron of a muon missing the sample) will be directly affected as follows. After an $\text{F}_{\text {center}}$ event, say a muon missing the sample (labeled $\text{M}_1$ and occurring at time $t_1$ -- note that this first event could also be a positron), a dead-time of $t_{\text{d}} \simeq 25$~ns occurs. Therefore, if a second muon (labeled $\text{M}_2$ and occurring at time $t_2$) misses the sample and ends-up in the $\text{F}_{\text {center}}$ within this dead-time (i.e. $t_1 \leq t_2 \leq t_1+t_{\text{d}}$), $\text{F}_{\text {center}}$ will not fire. As a consequence the event $\text{M}_2$ will mistakenly be considered as a good event as the condition $\text{M}\cdot\overline{\text{V}}$ will actually be fulfilled. 
In addition a consequent ``bad'' positron event, 
say $\text{P}_{\text {x}}$,  originating from the $\text{F}_{\text {center}}$ (e.g. the decay positron from either $\text{M}_1$ or $\text{M}_2$)  will still not be rejected by the $\text{F}_{\text {center}}$ during the period of time between the $\text{M}_2$ event and the end of the dead-time window after the $\text{M}_1$ event (i.e. between $t_2$ and $t_1+t_{\text{d}}$, see Fig~\ref{APD_deadtime}). Therefore, if P$_x$ is detected by a positron detector, a small but finite probability will exist that such pseudo-valid short-time $\text{M}_2$--$\text{P}_{\text {x}}$ events occur. 
To cure this distortion (see lower panel of Fig.~\ref{deadtime}), we implemented an asymmetric coincidence time-window for the definition of an M event, i.e. for the coincidence $\text{M}\cdot\overline{\text{V}}$. Hence, a muon event is defined as an M event without veto event in an asymmetric time-window extending from 25~ns before the M--event until a time $-t_{\text{ac}}/2$ (see Section~~\ref{Vetos_Anticoincidences}) after it.

 \subsubsection{Overall Time Resolution, Counting Rate and Magnetic Field Effects}
\label{alpha_in field}
\begin{center}
\begin{figure}[t]
\includegraphics[width=0.8\columnwidth]{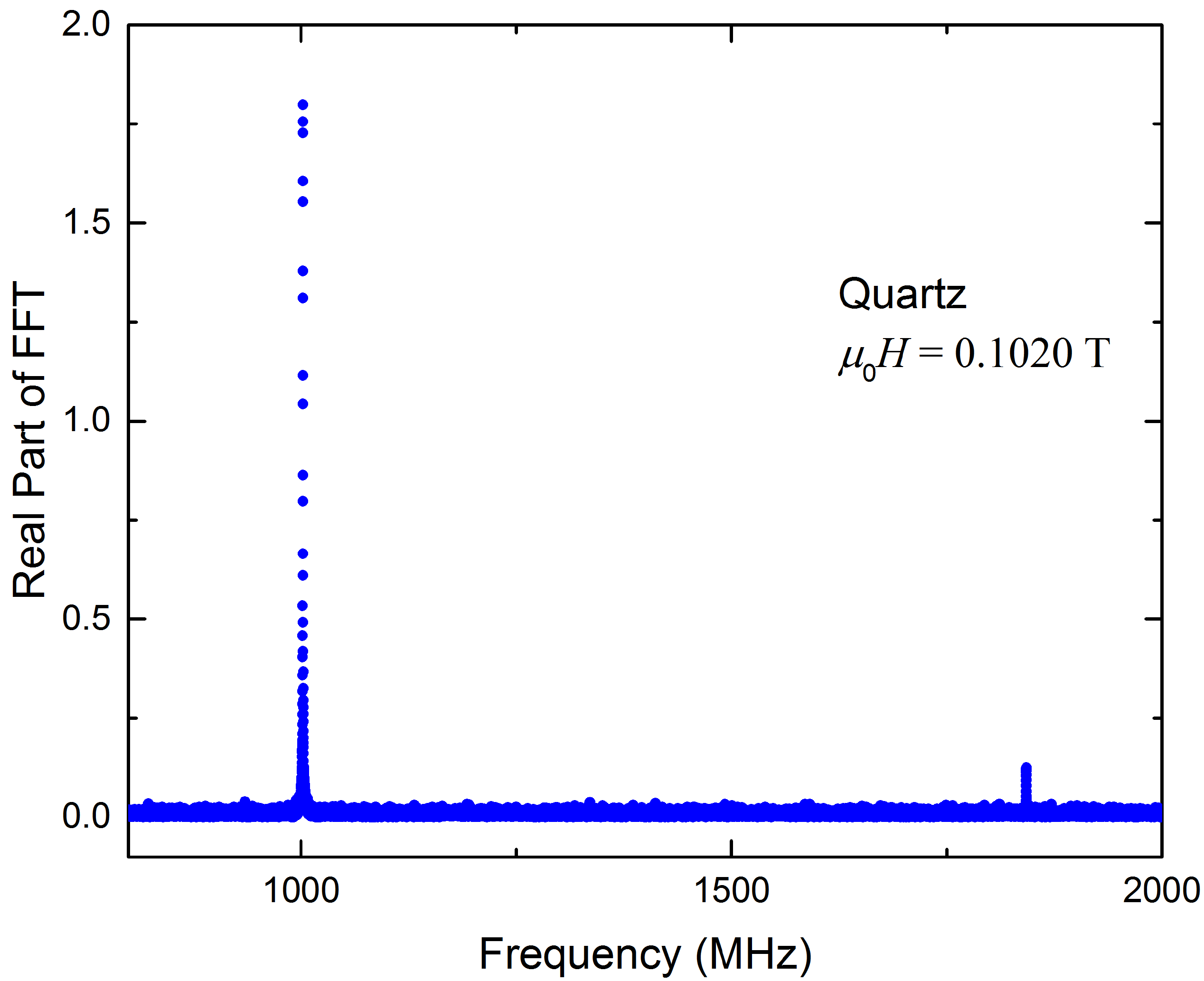}
\caption{Example of the frequency response of a quartz sample measured with an applied field of $\mu_0H=0.1020$~T. The observed frequencies at about 1002 and 1842~MHz correspond to intra-triplet precession frequencies of muonium. Note that the frequency arising from muons in the diamagnetic state (i.e. $\nu_{\mu}=13.82$~MHz) is not shown.}
\label{Quartz_FFT}
\end{figure}
\end{center}

The overall time resolution of the detector system has been obtained by measuring the amplitude change of the oscillatory component of the $\mu$SR signal recorded in a quartz sample ($15\times 15\times 5$~mm$^3$ synthetic sample). In quartz, a large fraction (ca 75\% in our case) of the muons pick up an electron and form muonium. In muonium the spins of the electron and the muon are coupled by the hyperfine interaction. In zero applied magnetic field, the four energy eigenvalues are grouped into two energy levels with different muonium total spin values $J$, a 
triplet state ($J=1$) and a singlet state ($J=0$). An applied transverse magnetic field will change the energy levels and will split the triplet state. As a result, and in addition to the precession of muons in diamagnetic states, the $\mu$SR spectrum will present precession frequencies corresponding to the splitting of the muonium energy levels\cite{Hughes}. At low applied field, the two lowest muonium frequencies can be determined by the GPS instrument and correspond to the intra-triplet precession frequencies of muonium (see Fig.~\ref{Quartz_FFT}).
The amplitude of the muonium signals is not only determined by the polarization associated with the corresponding precessions but also by the finite time resolution of the detector system\cite{Holzschuh}.

Correcting, for each signal of frequency $\nu_i$, the fitted apparent asymmetry $A_{\text {Mu,}i\text{,app}}$ 
for the corresponding calculated polarization $P_{\text {Mu,}i}$ one obtains the apparent full asymmetry for the muonium state $A_{\text {Mu,tot,app}}$:
\begin{equation}
 \frac{A_{\text {Mu,}i\text{,app}}(\nu_i,H)}{P_{\text {Mu,}i}(H)}=A_{\text {Mu,tot,app}}(\nu_i,H)
\end{equation}
 Note that $P_{\text {Mu,}i}$ is calculated by neglecting the small uniaxial anisotropy present at room temperature.
This apparent full asymmetry is reduced with respect to its true value $A_{\text {Mu,tot}}$ due to the time resolution with standard deviation $\sigma$: 
\begin{equation}
A_{\text {Mu,tot,app}}(\nu_i,H)= A_{\text {Mu,tot}}\,P_{\text {Mu,}i}(H)\exp[-2(\pi\sigma\nu_i)^2]
\label{Eq_Aapp}
\end{equation}
\begin{center}
\begin{figure}[t]
\includegraphics[width=0.8\columnwidth]{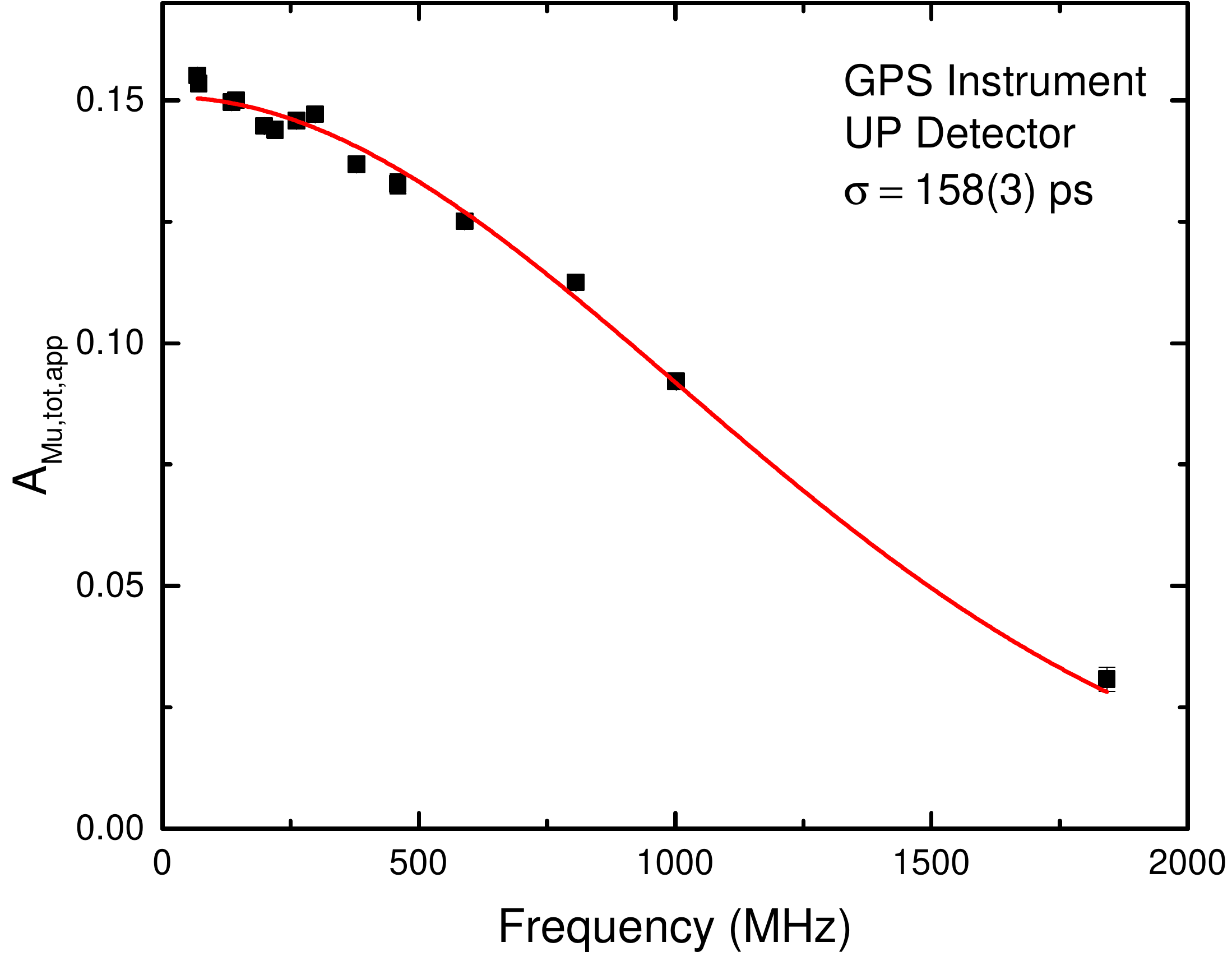}
\caption{Magnetic field evolution of the apparent asymmetry of the muonium signal (parameter $A_{\text {Mu,tot,app}}$, see text). 
The red line represents a fit using Eq.~\ref{Eq_Aapp}.
}
\label{fig_Aapp}
\end{figure}
\end{center}

Figure~\ref{fig_Aapp} exhibits the observed decrease of $A_{\text {Mu,tot,app}}$ observed with the Up detector together with the best fit of Eq.~\ref{Eq_Aapp} to the data obtained with a time resolution $\sigma= 158(3)$~ps. Note that identical results are obtained using the Down detector. Such a time resolution represents an improvement by more than a factor 5 with respect to the old GPS spectrometer which was equipped with standard photomultipliers and light guides.

By choosing a time-window of 10~$\mu$s, and therefore limiting the rate of stopped muons to about 35~kHz to restrict the number of pile-up events, one obtains overall count rates of validated muon-positron events of $4.0\times10^{7}/\text{hour}$ and $3.2\times10^{7}/\text{hour}$ using the ``No-Veto'' and ``Veto'' modes, respectively. Note that the count rate in ``Veto'' mode is reachable even in samples with reduced lateral dimensions (i.e. $\lesssim 10~\text{mm}^2$) without affecting the value of the asymmetry parameter. If the physics to be observed allows one to decrease the time-window, the event rate can be increased accordingly. For example, by choosing a time-window of 2.5~$\mu$s an event rate of 
 about $8.0\times10^{7}/\text{hour}$ can be reached using the ``No-Veto'' mode \cite{Note_Poisson}.
 
\begin{center}
\begin{figure}[t]
\includegraphics[width=0.85\columnwidth]{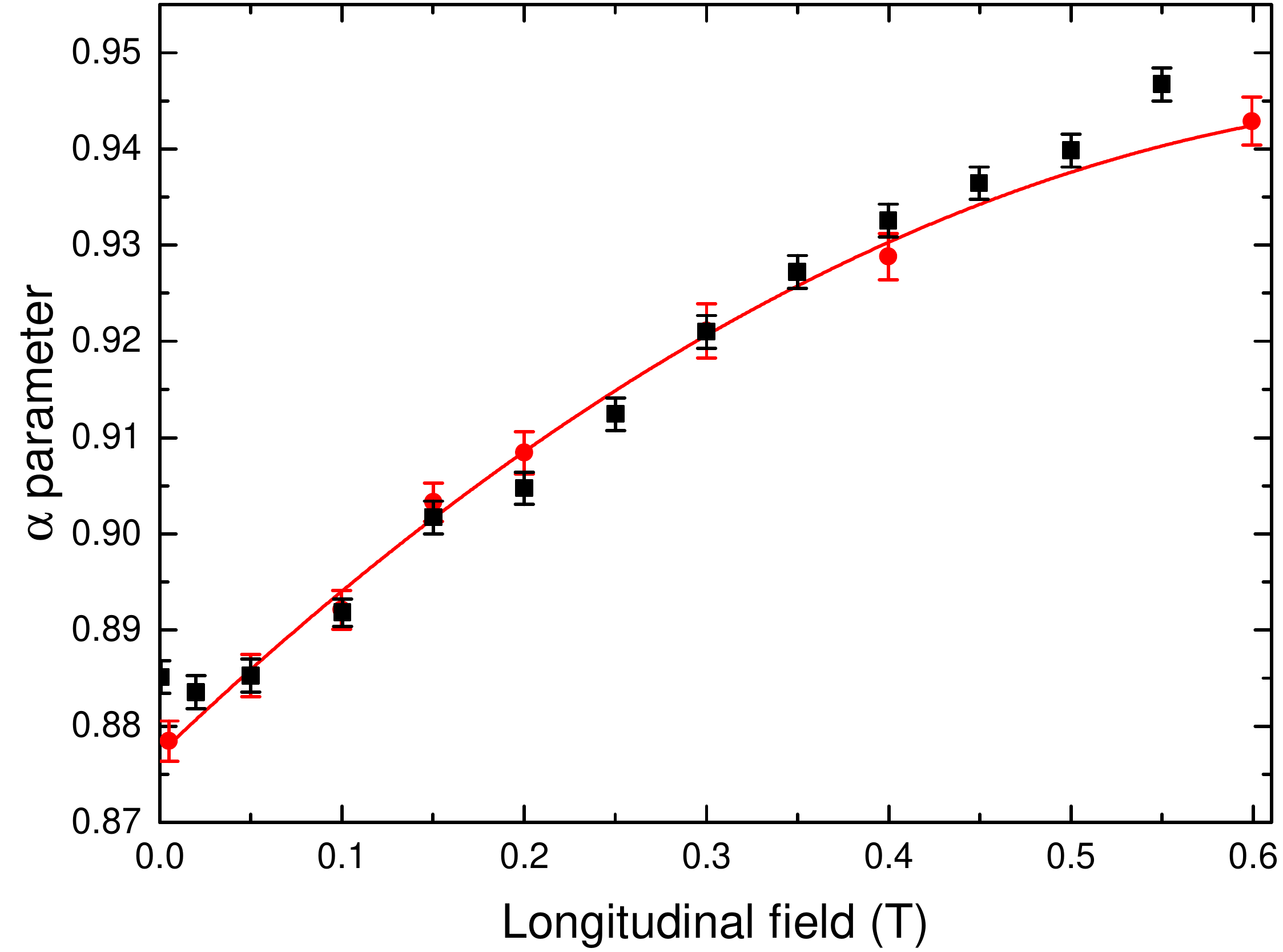}
\caption{ Field-evolution of the $\alpha$ parameter deduced from the Forward and Backward
detectors by assuming a constant asymmetry parameter $A$. Black symbols: measurements performed in
``Veto" mode on a $4\times 4\text{~mm}^2$ silver sample. Red symbols: results from musrSim simulations. The line is to guide the eye and represents a polynomial (2nd order) fit of the simulations.
 }
\label{figure_alpha_shift}
\end{figure}
\end{center}

As already reported, the applied magnetic field (here reaching values up to 0.65~T) is found to have no effect on the performances of the SiPM-based detectors. However, as the dimensions of the positron detectors are of course finite, a known effect of the applied field is to modify the value of the $\alpha$ parameter (see Eq.~\ref{asy_formula}) by altering the trajectories of the detected positrons. Figure~\ref{figure_alpha_shift} represents the field-induced shift of the $\alpha$ parameter as determined experimentally and by simulations in the so-called longitudinal-field (LF) configuration analyzing the Forward and Backward histograms. Note that these detectors, even though having similar solid angles, have slightly different shapes and are not precisely located at the same distance from the sample. 

\begin{center}
\begin{figure}[b]
\includegraphics[width=0.85\columnwidth]{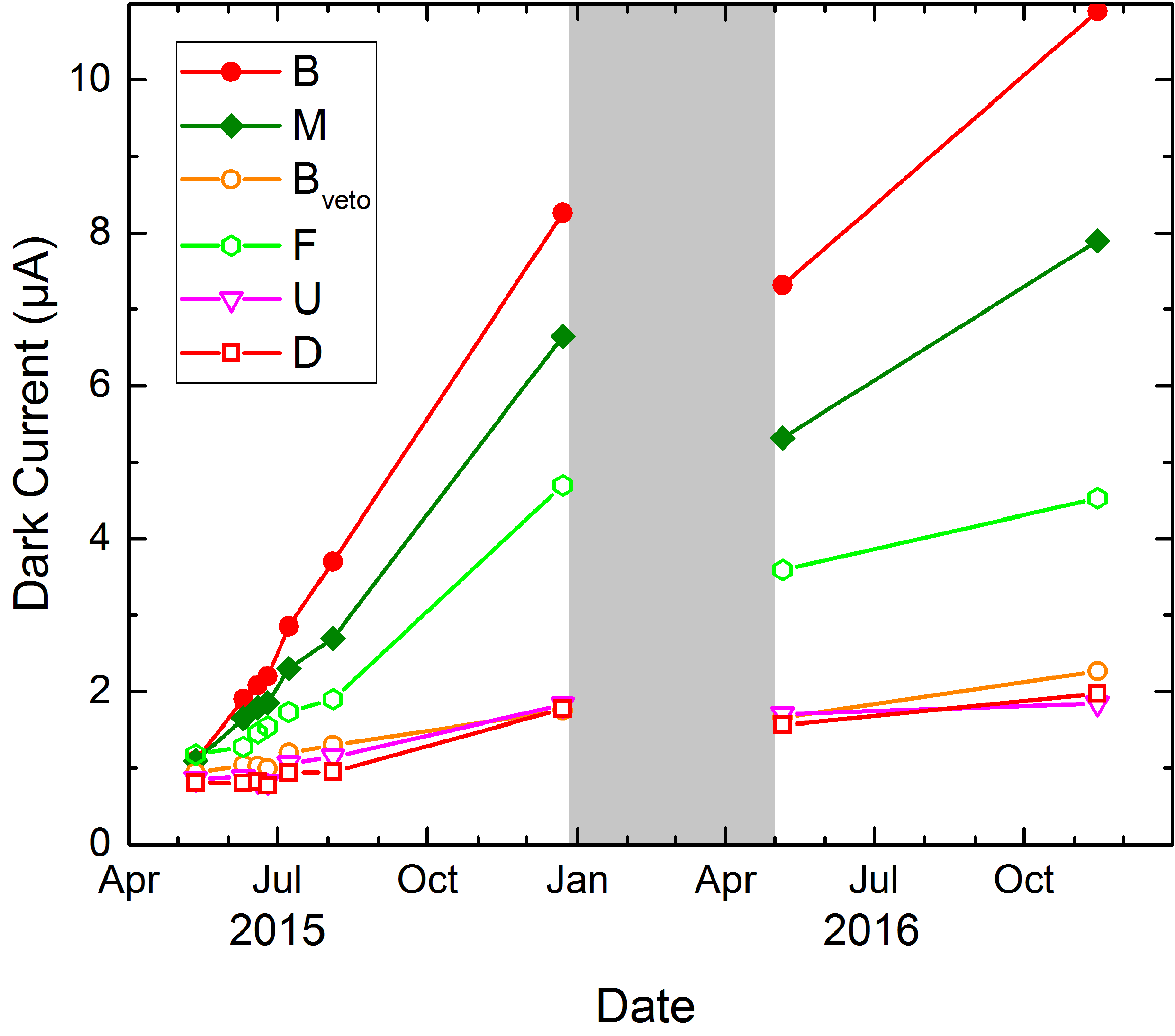}
\caption{Time-evolution of the dark current measured on different banks of SiPMs reading one side of the indicated detectors (beamtime operation for the years 2015-2016). The dark currents recorded for the opposite banks have very similar values. Note the increase is more pronounced for detectors close to the incoming beam and sample position (see Figure~\ref{figure_detectors}). The gray area indicates the interruption of the beam due to the usual maintenance shutdown of the HIPA complex. Note the slight but consistent decrease of the dark current values recorded prior and after the shutdown (see also text).
 }
\label{figure_dark_current}
\end{figure}
\end{center}

Another factor possibly affecting the SiPMs is the damage arising from radiation and evidenced by an increased dark current (see Fig.~\ref{figure_dark_current}). The main source of radiation damage arises from the decay positrons as no detector readout is located in the incoming beam. These decay positrons originate not only from the sample but actually mainly from the region near the center of the backward ``pyramid'' veto $\text{B}_{\text {veto}}$ (see Fig.~\ref{figure_detectors}). Figure~\ref{figure_dark_current} clearly indicates that the observed increase of the dark currents is more pronounced on SiPMs with an area subtending a larger solid angle with respect to this positron source. Note that the SiPMs for the $\text{B}_{\text {veto}}$ detector, whose center represents the main source of positrons, are themselves located rather far away from the center region leading to a restricted solid angle subtended by the SiPMs area respect to the positron source and consequently to a very slight dark current increase. The positron radiation causes primarily ionization damage in the surface region of the SiPMs and displacement damage, however, are thought to occur on a very small scale compared to protons for example \cite{Kataoka, Kodet, Musienko}. Note also the annealing effect occuring within the shutdown period. The observed increase of the SiPM dark current is found to produce no significant effects on the quality and rate of the data taking.

\section{Sample Environment}%
The available sample environment is similar to the one which was in use with the old GPS instrument. Two cryogenics ports are available on the instrument (see Fig.~\ref{New_GPS_view}). The first port is permanently equipped with a dynamic continuous $^4$He flow cryostat. This port shares the same isolation vacuum as the GPS instrument itself, i.e. as the muon beam and detectors, to minimize the number of titanium windows to be passed by the muon beam. The cryostat is based on Quantum Technology design which has been extensively retrofitted and upgraded in-house to achieve stable and reproducible operation. It is equipped with a phase separator from where the liquid helium is transported via a capillary to the sample chamber. Continuous operation down to 1.6~K is possible. At the level of the muon beam axis, the sample chamber is equipped with 10~$\mu$m thick titanium windows located up- and down-stream along the beam axis with respect to the sample position. A sample change takes about 5 min and the lowest temperature can be reached within 10 min after a sample change.

The second cryogenics port has an isolation vacuum separated from the one of the GPS instrument itself. On this port an oven with the sample in vacuum can be operated up to 1200 K. Alternatively, a horizontal ``side-loading" closed-cycle refrigerator cryostat allows one to cover low and high temperature ranges. The second stage of a Gifford-McMahon cold head of Sumitomo Heavy Industries, with a cooling power of 1 W at 4.2 K, is used to cool the end part of the chamber (i.e. tube) in which the sample stick is slided. This chamber can be operated either filled with helium gas (static configuration) for the low temperature regime down to 4 K, or can be evacuated (``warm-finger" configuration) for temperatures up to about 500 K. 

A configuration change between both cryogenics ports takes about 5 min, allowing a high turnover.

\section{Magnetic Fields}
  \subsection{Experimental Magnets}
The field provided by the main magnet see Fig~\ref{New_GPS_view}) is used either for LF or TF $\mu$SR experiments. For this latter type of experiments, the muon-spin is rotated with respect to the muon linear momentum by the spin-rotator (i.e. Wien filter) located about 10~m upstream in the beamline. As the field produced by the main magnet is applied along the muon linear momentum, no Lorentz force acts on the muon trajectory. The muon-spin rotation in the present spin-rotator represents about $50^\circ$.  
The conventional copper Helmholtz coils of the new main Magnet are operated with a 800~A/200~V power supply designed and produced at PSI. The maximum field is 0.78~T. Note that this corresponds to an increase of about 25\% compared to the old GPS instrument, which was made possible by using a double-conductor configuration for each sector of the coils. The field homogeneity in a $10\times 10\times 1\text{~mm}^3$ volume at the sample position (smaller dimension along the beam axis) is of the order of 1~ppm. The coils of the auxiliary magnet deliver up to 12~mT and are mainly used to determine the $\alpha$ parameter for ZF experiments.
  \subsection{Field Compensation}
Muon spin spectroscopy measurements rely on the proper shielding of the sample region from unwanted
magnetic fields originating e.g. from the earth's magnetic field or other external field sources within the 
experimental hall. In the past, this shielding has been achieved for the $\mu$SR spectrometers at PSI by three
pairs of orthogonal coils mounted around the sample position which were manually set to a fixed current
to compensate for static magnetic fields. The currents through the compensation coils were calibrated on a
regular basis every few weeks to maintain zero-field conditions at the sample. With this procedure a field
stability of $\pm 3$~$\mu$T has been obtained. Recently at PSI different particle physics experiments located in the same experimental hall as the GPS instrument have been equipped with large solenoid magnets. The fringe fields of these magnets
at the GPS instrument position adds up to a level of 0.2~mT. Since stray fields of this order of magnitude severely disturb the $\mu$SR experiments, especially if these fields are time-dependent, an active zero-field compensation device has been developed. 

The new zero-field compensation system utilizes the instrument compensation coils with the corresponding currents being dynamically adjusted to account for the time-dependent disturbing fields. To measure these fields, a 3-axis fluxgate probe
(Mag-03MC from Bartington Instruments) is permanently mounted near to the sample position. The three analog
outputs of the field probe are connected to a data acquisition and control unit. This unit communicates with the $\mu$SR slow control system via the MIDAS slow control bus (MSCB). A MIDAS frontend program logs the measured field values and calculates the desired compensation currents which are eventually provided by the control unit. As the position of the 3-axis fluxgate probe is not exactly at the sample position, a careful correction has to be applied to the calculation of the compensation currents to take into account the specific orientation of the probe, the difference between the values of the weak magnetic remanence determined at the probe and sample positions, and the difference between the values of magnetic field components created by the compensation coils pairs at both positions. With this system an active zero-field compensation with a short and long-term 
stability of $\pm 1$~$\mu$T is obtained.
\begin{center}
\begin{figure}[tb]
\includegraphics[width=0.85\columnwidth]{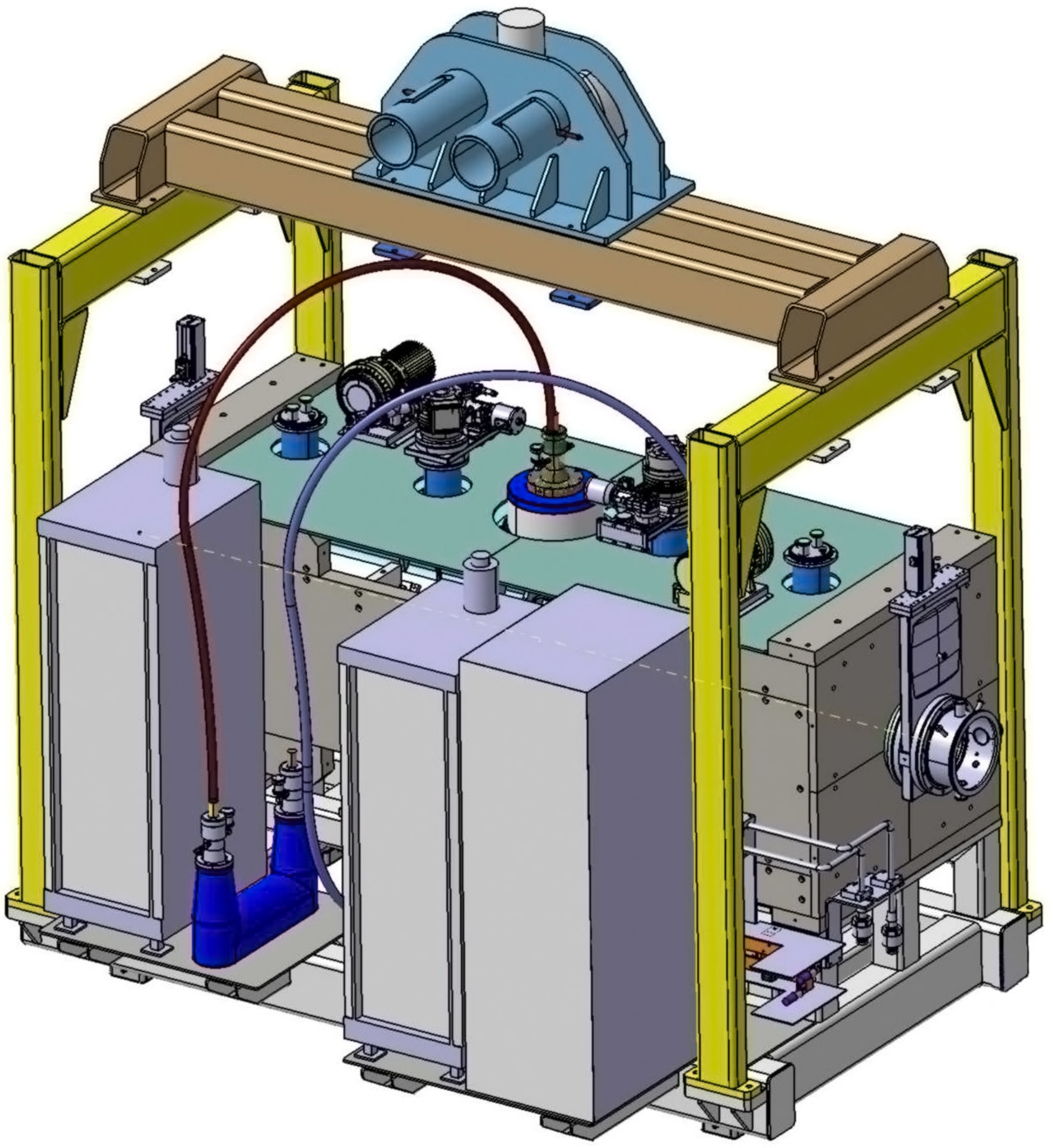}
\caption{Schematics of the new spin-rotator device located on the $\pi$M3.1 beamline of the HIPA complex. The spin rotation is performed in the vertical plane by horizontal magnets and the trajectory is corrected by electrodes sustaining a voltage of $\pm300$~kV.
 }
\label{figure_spin_rotator}
\end{figure}
\end{center}
\section{Future Projects and Conclusions}
Different upgrade projects are underway. A complete refurbishment of the ``Muon On REquest (MORE)" \cite{abela} fast switching deflector has been initiated and should be operational in 2018. Though the main characteristics of the MORE deflector equipment will remain unchanged, this upgrade will ensure a stable operation. Another major upgrade will take place with the replacement of the muon spin-rotator (see Fig.~\ref{figure_spin_rotator}). The new device has been fully developed and designed at PSI (see also Ref.~\onlinecite{Deiters,Vrankovic}) and is being constructed in collaboration with the Swiss industry. With this spin-rotator the rotation of the muon spin with respect to the muon linear momentum can reach up to $75^\circ$ which will lead to an increase of more than 25\% of the asymmetry of the $\mu$SR signal in TF configuration. The commissioning is foreseen at the end of 2017.

In conclusion, the commissioning of the new state-of-the-art GPS $\mu$SR instrument has been successful and entirely fulfilled  the expectations based on extended simulation studies. The core of the instrument is formed by particle detectors based on SiPM technology. The compactness of these detectors has been exploited to increase the solid angle covered by the positron detectors and led to a significant improvement of the time resolution. 
The instrument is in normal operation at S$\mu$S since June 2015.  
\vspace{3mm}~

\noindent\rule[0.5ex]{\linewidth}{1pt}

\begin{thebibliography}{99}
\bibitem{Note_Poisson} As the muon events are Poisson distributed, note that to maintain the pile-up events at the same level as in the case of a 10~$\mu\text{s}$ time-window, an incoming-muon rate of 140~kHz is in principle allowed with a 2.5~$\mu\text{s}$ time-window. In reality a reduction to about 100~kHz is necessary to maintain the veto-event rate in the pyramid below about 1~MHz, which represents the limit per channel for the TDC.
\end{thebibliography}
\end{document}